# Coupled cavity model based on the mode matching technique

M.I. Ayzatsky[1], V.V.Mytrochenko

National Science Center Kharkov Institute of Physics and Technology (NSC KIPT),
610108, Kharkov, Ukraine

We have developed the mode matching technique that is based on the using the eigenmodes of circular cavities and the eigenwaves of circular waveguides as the basic functions for calculation the properties of nonuniform DLW. We have obtained exact infinite systems of coupled equations which can be reduced by making some assumptions. Under such procedure we can receive more exact parameters of nonuniform equivalent circuits by solving the appropriative algebraic systems. These parameters of equivalent circuits are functions both geometric sizes and frequency. Moreover, under such approach all used values have interpretation. We called this approach as coupled cavity model.

## 1 Introduction

Disc-loaded waveguides (DLW) have been heavily investigated both numerically and analytically over the past seven decades (see, for example, [1,2] and cited there literature). They have also been used, and continue to be used, in a variety of microwave devices such as linear accelerators [3,4], travelling-wave tube amplifiers, backward-wave oscillators [5], etc.

There are several known reliable methods and computer codes for calculations of the infinitely periodic structures with using Floquet theorem and discretizing Maxwell's equations.

Usually the electromagnetic properties of accelerator components are calculated by computer codes that discretize Maxwell's equations. For long tapered disc-loaded waveguides, however, these methods would need the solution of extremely large algebraic equations. This is numerically difficult (or even impossible).

So, it is necessary to use non-grid-oriented methods to calculate the fields in the complete structure with realistic dimensions. Equivalent circuits are one possible technique. These are fast methods but the influence of the chosen model on the results is not negligible so that the results may be far away from the exact solution of Maxwell's equations. The mode matching technique is based on an exact formulation. In the chain matrix formulation this technique can be used for periodic and nonperiodic structures [6,7,8,9].

Usually, in the mode matching technique basic functions are chosen as the eigenwaves of circular waveguides [6,8,9]. Earlier we have developed approach that used the eigenmodes of circular cavities as the basic functions for calculation the properties of uniform DLW [10]. We have obtained exact infinite system of coupled equations which can be reduced by making some assumptions. Under such procedure we can receive more exact parameters of equivalent circuits by solving the appropriative algebraic systems. These parameters of equivalent circuits are functions both geometric sizes and frequency. Moreover, under such approach all used values have interpretation. We called this approach as coupled cavity model. Here we present the extension of that model on the case of nonuniform DLW.

## 2 Electromagnetic fields in nonuniform disk-loaded waveguide

Let's consider a cylindrical nonuniform DLW (Fig.1). We will consider only axially symmetric fields with $E_z, E_r, H_\varphi$ components. Time dependence is $\exp(-i\omega t)$.

---

[1] M.I. Aizatskyi, N.I.Aizatsky; aizatsky@kipt.kharkov.ua

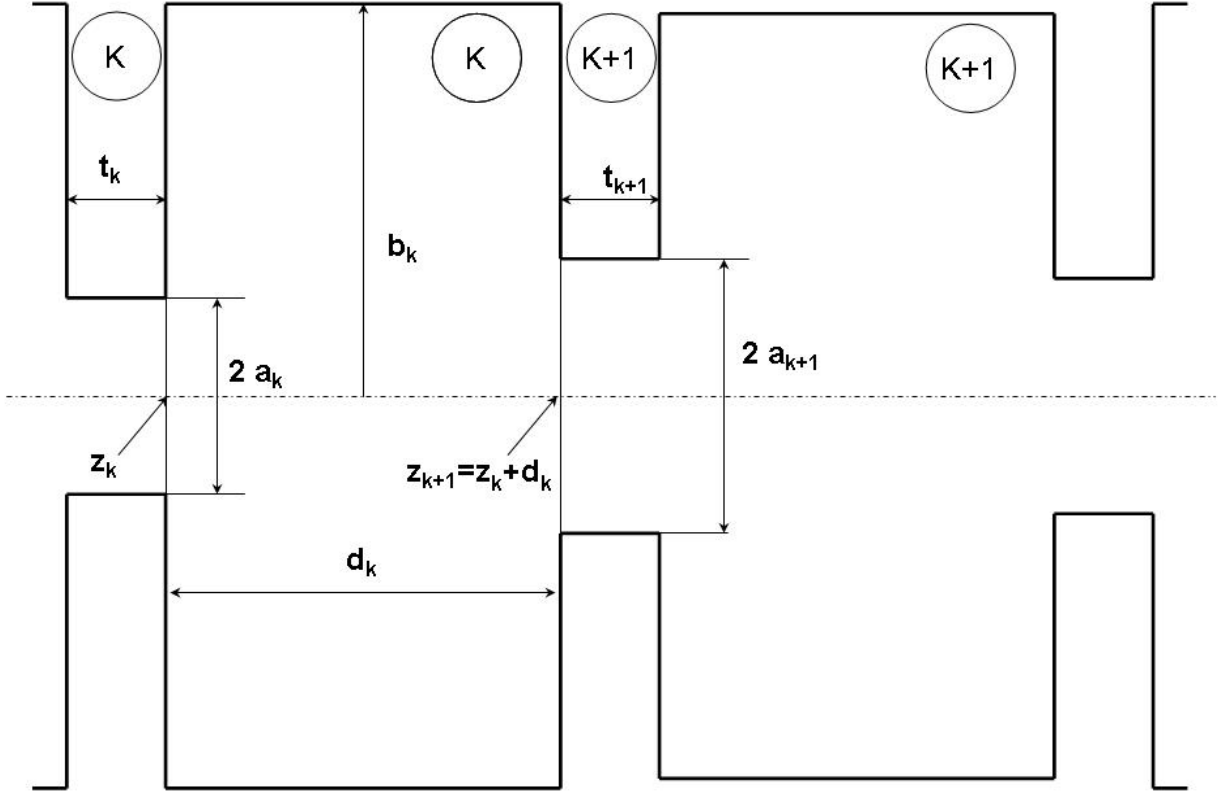

**Fig. 1**

We can divide the DLW volume into infinite number of different cylindrical volumes that are contiguous with each other over circle aria. In each large volume we expand the electromagnetic field with the short-circuit resonant cavity modes

$$\vec{E}^{(k)} = \sum_q e_q^{(k)}(t)\vec{E}_q^{(k)}(\vec{r}) , \qquad (1.1)$$

$$\vec{H}^{(k)} = i\sum_q h_q^{(k)}(t)\vec{H}_q^{(k)}(\vec{r}), \qquad (1.2)$$

where $q = \{0, m, n\}$.

$\vec{E}_q^{(k)}$, $\vec{H}_q^{(k)}$ are the solutions of homogenous Maxwell equations

$$\begin{aligned} rot\, \vec{E}_q^{(k)} &= i\omega_q^{(k)}\mu_0\vec{H}_q^{(k)} ,\\ rot\, \vec{H}_q^{(k)} &= -i\omega_q^{(k)}\varepsilon_0\vec{E}_q^{(k)} \end{aligned} \qquad (1.3)$$

with boundary condition $\vec{E}_\tau = 0$ on the metal surface

$$E_{m,n,z}^{(k)} = J_0\left(\frac{\lambda_m}{b_k}r\right)\cos\left(\frac{\pi}{d_k}n(z-z_k)\right), \qquad (1.4)$$

$$H_{m,n,\varphi}^{(k)} = -i\omega_{n,m}^{(k)}\frac{\varepsilon_0 b_k}{\lambda_m}J_1\left(\frac{\lambda_m}{b_k}r\right)\cos\left(\frac{\pi}{d_k}n(z-z_k)\right), \qquad (1.5)$$

$$E_{m,n,r}^{(k)} = \frac{b_k}{\lambda_m}\frac{\pi n}{d_k}J_1\left(\frac{\lambda_m}{b_k}r\right)\sin\left(\frac{\pi}{d_k}n(z-z_k)\right), \qquad (1.6)$$

$$\omega_{m,n}^{(k)2} = c^2 \left\{ \left(\frac{\lambda_m}{b_k}\right)^2 + \left(\frac{\pi n}{d_k}\right)^2 \right\}, \tag{1.7}$$

where $J_0(\lambda_m) = 0$.

In each small volume we expand the electromagnetic field with the waveguide modes

$$\vec{H}^{(k)} = \sum_s \left( C_s^{(k)} \vec{\mathcal{H}}_s^{(k)} + C_{-s}^{(k)} \vec{\mathcal{H}}_{-s}^{(k)} \right), \tag{1.8}$$

$$\vec{E}^{(k)} = \sum_s \left( C_s^{(k)} \vec{\mathcal{E}}_s^{(k)} + C_{-s}^{(k)} \vec{\mathcal{E}}_{-s}^{(k)} \right), \tag{1.9}$$

where

$$\mathcal{E}_{\pm s,z}^{(k)} = J_0\left(\frac{\lambda_s}{a_k} r\right) \exp\{\pm \gamma_s^{(k)}(z - z_k + t_k)\}, \tag{1.10}$$

$$\mathcal{H}_{\pm s,\varphi}^{(k)} = -i\omega \frac{\varepsilon_0 \varepsilon a_k}{\lambda_s} J_1\left(\frac{\lambda_s}{a_k} r\right) \exp\{\pm \gamma_s^{(k)}(z - z_k + t_k)\}, \tag{1.11}$$

$$\mathcal{E}_{\pm s,r}^{(k)} = \mp \frac{a_k}{\lambda_s} \gamma_s^{(k)} J_1\left(\frac{\lambda_s}{a_k} r\right) \exp\{\pm \gamma_s^{(k)}(z - z_k + t_k)\}, \tag{1.12}$$

$$\gamma_s^{(k)2} = \left(\frac{\lambda_s}{a_k}\right)^2 - \frac{\omega^2}{c^2} = \frac{1}{a_k^2}\left(\lambda_s^2 - \frac{a_k^2 \omega^2}{c^2}\right). \tag{1.13}$$

Lets us chose $E_{010}$ modes of the cavities as the basic modes (oscillators) [10,11]. Then using the relevant boundary conditions after some manipulations we can obtain an infinite set of infinite systems of linear equations for coefficients that determine fields in apertures

$$\tilde{C}_{l,s}^{(k)} + \frac{1}{ch(\gamma_s^{(k)} t_k)} \sum_{s'} \tilde{C}_{r,s'}^{(k)} T_{s,s'}^{(k,1)} + \sum_{s'} \tilde{C}_{l,s'}^{(k)} T_{s,s'}^{(k,2)} - \frac{1}{ch(\gamma_s^{(k)} t_k)} \sum_{s'} \tilde{C}_{l,s'}^{(k+1)} T_{s,s'}^{(k,3)} -$$
$$- \sum_{s'} \tilde{C}_{r,s'}^{(k-1)} T_{s,s'}^{(k,4)} = R_s^{(k,1)} ch^{-1}\left(\gamma_s^{(k)} t_k\right) e_{0,1}^{(k)} - R_s^{(k,2)} e_{0,1}^{(k-1)} = \sum_j \delta_{j,k} \left\{ ch^{-1}\left(\gamma_s^{(k)} t_k\right) R_s^{(k,1)} e_{0,1}^{(j)} - R_s^{(k,2)} e_{0,1}^{(j-1)} \right\}$$
$$\tilde{C}_{r,s}^{(k)} + \sum_{s'} \tilde{C}_{r,s'}^{(k)} T_{s,s'}^{(k,1)} + \frac{1}{ch(\gamma_s^{(k)} t_k)} \sum_{s'} \tilde{C}_{l,s'}^{(k)} T_{s,s'}^{(k,2)} - \sum_{s'} \tilde{C}_{l,s'}^{(k+1)} T_{s,s'}^{(k,3)} - \frac{1}{ch(\gamma_s^{(k)} t_k)} \sum_{s'} \tilde{C}_{r,s'}^{(k-1)} T_{s,s'}^{(k,4)} =$$
$$= R_s^{(k,1)} e_{0,1}^{(k)} - R_s^{(k,2)} ch^{-1}\left(\gamma_s^{(k)} t_k\right) e_{0,1}^{(k-1)} = \sum_j \delta_{j,k} \left\{ R_s^{(k,1)} e_{0,1}^{(j)} - ch^{-1}\left(\gamma_s^{(k)} t_k\right) R_s^{(k,2)} e_{0,1}^{(j-1)} \right\}$$
,(1.14)

where

$$\tilde{C}_{l,s}^{(k)} = \frac{\overline{\gamma}_s^{(k)}}{\lambda_s} J_1(\lambda_s) \left[ -C_s^{(k)} + C_{-s}^{(k)} \right] \tag{1.15}$$

$$\tilde{C}_{r,s}^{(k)} = \frac{\overline{\gamma}_s^{(k)}}{\lambda_s} J_1(\lambda_s) \left[ -C_s^{(k)} \exp\left(\gamma_s^{(k)} t_k\right) + C_{-s}^{(k)} \exp\left(-\gamma_s^{(k)} t_k\right) \right] \tag{1.16}$$

$$T_{s,s'}^{(k,1)} = \frac{\bar{\gamma}_s^{(k)} 4 a_k^3}{b_k^3} \frac{1+\exp(-2\gamma_s^{(k)}t_k)}{1-\exp(-2\gamma_s^{(k)}t_k)} \sum_m \Lambda_m^{(1,k)} \frac{\lambda_m^2}{J_1^2(\lambda_m)} \frac{J_0^2\left(\frac{a_k \lambda_m}{b_k}\right)}{\left(\frac{a_k \lambda_m}{b_k}\right)^2 - (\lambda_s)^2} \frac{1}{\left(\frac{a_k \lambda_m}{b_k}\right)^2 - (\lambda_{s'})^2}$$

$$T_{s,s'}^{(k,2)} = \frac{\bar{\gamma}_s^{(k)} 4 a_k^3}{b_{k-1}^3} \frac{1+\exp(-2\gamma_s^{(k)}t_k)}{1-\exp(-2\gamma_s^{(k)}t_k)} \sum_m \Lambda_m^{(1,k-1)} \frac{\lambda_m^2}{J_1^2(\lambda_m)} \frac{J_0^2\left(\frac{a_k \lambda_m}{b_{k-1}}\right)}{\left(\frac{a_k \lambda_m}{b_{k-1}}\right)^2 - (\lambda_s)^2} \frac{1}{\left(\frac{a_k \lambda_m}{b_{k-1}}\right)^2 - (\lambda_{s'})^2}$$

$$T_{s,s'}^{(k,3)} = \frac{\bar{\gamma}_s^{(k)} 4 a_{k+1}^3}{b_k^3} \frac{1+\exp(-2\gamma_s^{(k)}t_k)}{1-\exp(-2\gamma_s^{(k)}t_k)} \sum_m \Lambda_m^{(2,k)} \frac{\lambda_m^2}{J_1^2(\lambda_m)} \frac{J_0\left(\frac{a_k \lambda_m}{b_k}\right) J_0\left(\frac{a_{k+1} \lambda_m}{b_k}\right)}{\left(\frac{a_k \lambda_m}{b_k}\right)^2 - (\lambda_s)^2 \left(\frac{a_{k+1} \lambda_m}{b_k}\right)^2 - (\lambda_{s'})^2}$$

$$T_{s,s'}^{(k,4)} = \frac{\bar{\gamma}_s^{(k)} 4 a_{k-1}^3}{b_{k-1}^3} \frac{1+\exp(-2\gamma_s^{(k)}t_k)}{1-\exp(-2\gamma_s^{(k)}t_k)} \sum_m \Lambda_m^{(2,k-1)} \frac{\lambda_m^2}{J_1^2(\lambda_m)} \frac{J_0\left(\frac{a_k \lambda_m}{b_{k-1}}\right) J_0\left(\frac{a_{k-1} \lambda_m}{b_{k-1}}\right)}{\left(\frac{a_k \lambda_m}{b_{k-1}}\right)^2 - (\lambda_s)^2 \left(\frac{a_{k-1} \lambda_m}{b_{k-1}}\right)^2 - (\lambda_{s'})^2}, (1.17)$$

$$\Lambda_m^{(1,k)} = \begin{cases} \frac{1}{b_k h_1^{(k)}} \left[ cth(d_k h_1^{(k)}) - \frac{1}{d_k h_1^{(k)}} \right], & m=1 \\ \frac{1}{b_k h_m^{(k)}} cth(d_k h_m^{(k)}), & m>1 \end{cases}, \quad (1.18)$$

$$\Lambda_m^{(2,k)} = \begin{cases} \frac{1}{b_k h_1^{(k)}} \left[ \frac{1}{sh(d_k h_1^{(k)})} - \frac{1}{d_k h_1^{(k)}} \right], & m=1 \\ \frac{1}{b_k h_m^{(k)} sh(d_k h_m^{(k)})}, & m>1 \end{cases} \quad (1.19)$$

$$h_m^{(k)} = \sqrt{\frac{\lambda_m^2}{b_k^2} - \frac{\omega^2}{c^2}} \quad (1.20).$$

$$R_s^{(k,1)} = \frac{1+\exp(-2\gamma_s^{(k)}t_k)}{1-\exp(-2\gamma_s^{(k)}t_k)} \frac{2\bar{\gamma}_s^{(k)} J_0\left(\frac{a_k \lambda_1}{b_k}\right)}{\left(\frac{a_k \lambda_1}{b_k}\right)^2 - (\lambda_s)^2} -$$

$$R_s^{(k,2)} = \frac{1+\exp(-2\gamma_s^{(k)}t_k)}{1-\exp(-2\gamma_s^{(k)}t_k)} \frac{2\bar{\gamma}_s^{(k)} J_0\left(\frac{a_k \lambda_1}{b_{k-1}}\right)}{\left(\frac{a_k \lambda_1}{b_{k-1}}\right)^2 - (\lambda_s)^2} \quad , (1.21)$$

Mode amplitudes inside cavities can be found by summing the relevant series with the coefficients $\tilde{C}_{l,s}^{(k)}, \tilde{C}_{r,s}^{(k)}$

$$\left(\omega_{m,n}^{(k)2} - \omega^2\right)e_{m,n}^{(k)} = -\frac{\omega_{m,n}^{(k)2}\varepsilon_0 a_k}{N_{m,n}^{(k)}} 2\pi \sum_s \frac{J_0\left(\frac{a_k \lambda_m}{b_k}\right)}{\left(\frac{\lambda_m}{b_k}\right)^2 - \left(\frac{\lambda_s}{a_v}\right)^2}\tilde{C}_{r,s}^{(k)} +$$

$$+\frac{\omega_{m,n}^{(k)2}\varepsilon_0 a_{k+1}}{N_{m,n}^{(k)}} 2\pi(-1)^n \sum_s \frac{J_0\left(\frac{a_{k+1}\lambda_m}{b_k}\right)}{\left(\frac{\lambda_m}{b_k}\right)^2 - \left(\frac{\lambda_s}{a_{k+1}}\right)^2}\tilde{C}_{l,s}^{(k+1)}$$

(1.22)

where

$$N_{m,n}^{(k)} = \frac{\omega_{m,n}^{(k)2} b_k^2}{2c^2}\varepsilon_0 \pi d_k J_1^2(\lambda_m)\sigma_n,$$ (1.23)

$$\sigma_n = \begin{cases} 2 & n=0 \\ 1 & n>0 \end{cases}.$$ (1.24)

Solutions of the set of linear equations (1.14) can be expressed through amplitudes of $E_{010}$ modes

$$\tilde{C}_{l,s}^{(k+1)} = \sum_j \left(Y_{l,s}^{(1,k+1,j)} - Y_{l,s}^{(2,k+1,j)}\right)e_{01}^{(j)}$$

$$\tilde{C}_{r,s}^{(k)} = \sum_j \left(Y_{r,s}^{(1,k,j)} - Y_{r,s}^{(2,k,j)}\right)e_{01}^{(j)}$$

(1.25)

where new coefficients $Y$ do not depend on amplitudes of $E_{010}$ modes and are solutions of the new infinite set of infinite systems of linear equations

$$Y_{l,s}^{(1,k+1,j)} + \sum_{s'}Y_{l,s'}^{(1,k+1,j)}T_{s,s'}^{(k+1,2)} - \sum_{s'}Y_{r,s'}^{(1,k,j)}T_{s,s'}^{(k+1,4)} - \frac{1}{ch(\gamma_s^{(k+1)}t_{k+1})}\sum_{s'}Y_{l,s'}^{(1,k+2,j)}T_{s,s'}^{(k+1,3)} +$$

$$+\frac{1}{ch(\gamma_s^{(k+1)}t_{k+1})}\sum_{s'}Y_{r,s'}^{(1,k+1,j)}T_{s,s'}^{(k+1,1)} = R_s^{(k+1,1)}ch^{-1}\left(\gamma_s^{(k+1)}t_{i+1}\right)\delta_{k+1,j}$$

$$Y_{r,s}^{(1,k,j)} + \sum_{s'}Y_{r,s'}^{(1,k,j)}T_{s,s'}^{(k,1)} - \sum_{s'}Y_{l,s'}^{(1,k+1,j)}T_{s,s'}^{(k,3)} - \frac{1}{ch(\gamma_s^{(k)}t_k)}\sum_{s'}Y_{l,s'}^{(1,k,j)}T_{s,s'}^{(k,2)} +$$

$$+\frac{1}{ch(\gamma_s^{(k)}t_k)}\sum_{s'}Y_{r,s'}^{(1,k-1,j)}T_{s,s'}^{(k,4)} = R_s^{(k,1)}\delta_{k,j}$$

(1.26)

$$Y_{l,s}^{(2,k+1,j)} + \sum_{s'}Y_{l,s'}^{(2,k+1,j)}T_{s,s'}^{(k+1,2)} - \sum_{s'}Y_{r,s'}^{(2,k,j)}T_{s,s'}^{(k+1,4)} -$$

$$-\frac{1}{ch(\gamma_s^{(k+1)}t_{k+1})}\sum_{s'}Y_{l,s'}^{(2,k+2,j)}T_{s,s'}^{(k+1,3)} + \frac{1}{ch(\gamma_s^{(k+1)}t_{k+1})}\sum_{s'}Y_{r,s'}^{(2,k+1,j)}T_{s,s'}^{(k+1,1)} = R_s^{(k+1,2)}\delta_{k,j}$$

$$Y_{r,s}^{(2,k,j)} + \sum_{s'}Y_{r,s'}^{(2,k,j)}T_{s,s'}^{(k,1)} - \sum_{s'}Y_{l,s'}^{(2,k+1,j)}T_{s,s'}^{(k,3)} - \frac{1}{ch(\gamma_s^{(k)}t_k)}\sum_{s'}Y_{l,s'}^{(2,k-1,j)}T_{s,s'}^{(k,4)} +$$

$$+\frac{1}{ch(\gamma_s^{(k)}t_k)}\sum_{s'}Y_{r,s'}^{(2,k,j)}T_{s,s'}^{(k,2)} = R_s^{(k,2)}ch^{-1}\left(\gamma_s^{(k)}t_k\right)\delta_{k-1,j}$$

(1.27)

Using (1.25) we can rewrite (1.22) as

$$\left(\omega_{m,n}^{(k)2} - \omega^2\right)e_{m,n}^{(k)} = \omega_{m,n}^{(k)2}\sum_{j=-\infty}^{\infty}e_{1,0}^{(j)}\alpha_{m,n}^{(k,j)},$$ (1.28)

where

$$\alpha_{m,n}^{(k,j)} = \frac{2\pi \varepsilon_0 b_k^3}{N_{n,m}^{(k)} \lambda_m^3} \left[ -\sum_s \left(Y_{r,s}^{(1,k,j)} - Y_{r,s}^{(2,k,j)}\right) F_s\left(\frac{a_k}{b_k}\lambda_m\right) + \right.$$
$$\left. +(-1)^n \sum_s \left(Y_{l,s}^{(1,k+1,j)} - Y_{l,s}^{(2,k+1,j)}\right) F_s\left(\frac{a_{k+1}}{b_k}\lambda_m\right) \right]$$
(1.29)

$$F_s(x) = \frac{x^3 J_0(x)}{(x)^2 - (\lambda_s)^2}.$$
(1.30)

Let's note that relations (1.28) are the exact ones.

It follows from (1.28) that for finding the amplitudes of the main $E_{010}$ mode we have to solve a system of coupled equations

$$\left(\omega_{1,0}^{(k)2} - \omega^2\right) e_{1,0}^{(k)} = \omega_{1,0}^{(k)2} \sum_{j=-\infty}^{\infty} e_{1,0}^{(j)} \alpha_{1,0}^{(k,j)}.$$
(1.31)

Amplitudes of other modes ($(m,n) \neq (1,0)$) can be find by summing the relevant series

$$e_{m,n}^{(k)} = \frac{\omega_{m,n}^{(k)2}}{\left\{\omega_{m,n}^{(k)2}\left(1 - \alpha_{m,n}^{(k,k)}\right) - \omega^2\right\}} \sum_{j=-\infty}^{\infty} e_{1,0}^{(j)} \alpha_{m,n}^{(k,j)}.$$
(1.32)

The system of coupled equations (1.31) is very similar to that one that can be constructed on the base of equivalent circuits approach. But in our approach the coefficients $\alpha_{m,n}^{(v,j)}$ are electrodynamically strictly defined and can be calculated with necessary accuracy. They depend on both the frequency $\omega$ and geometrical sizes of all volumes.

As it follows from physics of couplings, the contribution of "long range" couplings is small[2] and one can confine oneself to consideration a finite number of adjacent couplings. In this case the system of coupled equations (1.31) transforms into such one

$$\left(\omega_{1,0}^{(k)2} - \omega^2\right) e_{1,0}^{(k)} = \omega_{1,0}^{(k)2} \sum_{j=k-N}^{k+N} e_{1,0}^{(j)} \alpha_{1,0}^{(k,j)}.$$
(1.33)

For dealing with this system we have to develop procedure for $\alpha_{1,0}^{(k,j)}$ calculation. From (1.29) it follows that we have to know the coefficients

$$\left(Y_{l,s}^{(1,k+1,j)}, Y_{r,s}^{(1,k,j)}\right), \left(Y_{l,s}^{(2,k+1,j)}, Y_{r,s}^{(2,k,j)}\right)$$
(1.34)

for $k - N \leq j \leq k + N$. Coefficients $Y$ are the solutions of the infinite set of infinite systems of linear equations (1.26) and (1.27). We suppose that this set can be truncated and solution of such truncated set of equations converge under increasing the number of equations. Besides, we have to make restriction on the number of waveguide modes in the field expansion inside apertures. Used truncation method for the case $N = 4$ is explained in Fig. 2. Coefficients $Y_{l,s}^{(1,i,j)}$ connect with the tangential electric field on the left side of aperture with number $i$, $Y_{r,s}^{(1,i,j)}$ - with the tangential electric field on the right side of that aperture. As circles (with suitable place with respect to the diaphragms) we depicted coefficients $Y$ that enter into the first subset in expressions (1.26) (it is also true for the coefficients $Y$ that enter into the second subset in expressions (1.26)). One can see that presented finite set of linear equations is closed (contains 18 systems for 18 groups of unknown values).

---

[2] In the case under consideration (when $E_{010}$ modes are the basic ones "long range" ) coupling is realized through evanescent fields

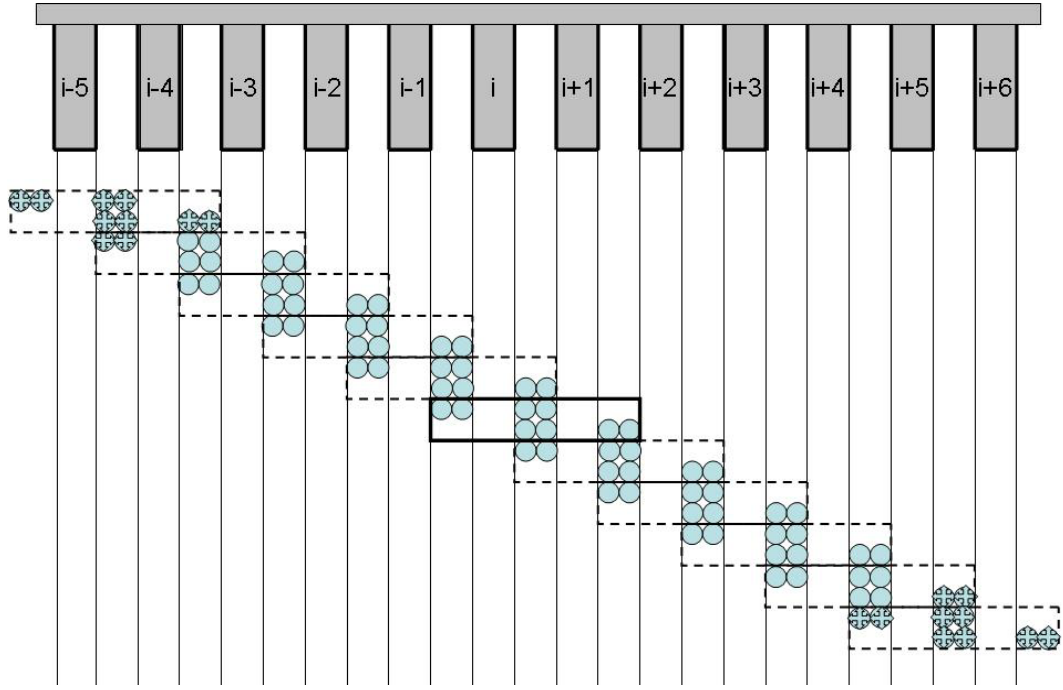

**Fig. 2 Truncation scheme for $N=4$**

It is necessary to introduce some criterion for choosing the value $N$ that can give correct simulation results. As such criterion we chose the accuracy of calculating of phase shift per cell for the case of homogeneous DLW. In this case amplitudes $e_{1,0}^{(k)}$ obey the Floquet theorem

$$e_{1,0}^{(k)} = e_{1,0}^{(0)} \exp(ik\varphi) \tag{1.35}$$

and the dispersive equation has such form [10]

$$\left\{\omega_{1,0}^{(0)2}\left(1+\alpha_{1,0}^{(0,0)}\right)-\omega^2\right\} = 2\omega_{1,0}^{(0)2}\sum_{j=1}^{N}\alpha_{1,0}^{(0,j)}(\omega)\cos(j\varphi). \tag{1.36}$$

For given $\varphi=\varphi_*, N$ and geometrical sizes we can find from this equation corresponding frequency $\omega=\omega_*$.

Then, for this frequency we can find the phase shift that satisfy equation

$$\left\{\omega_{1,0}^{(0)2}\left(1+\alpha_{1,0}^{(0,0)}\right)-\omega_*^2\right\} = 2\omega_{1,0}^{(0)2}\sum_{j=1}^{n}\alpha_{1,0}^{(0,j)}(\omega_*)\cos(j\varphi) ,. \tag{1.37}$$

where $n<N$.

As an example, in Table 1 we present values $\alpha_{1,0}^{(0,j)}$ for two types of homogeneous DLW.

**Table 1**

|  | $a=1$ cm, $b=4.075$ cm, $d==3.1$ cm | $a=1.8$ cm, $b=4.075$ cm, $d==1.55$ cm |
|---|---|---|
| $\alpha_{1,0}^{(0,-3)}$ | 1.75E-09 | 2.30E-04 |
| $\alpha_{1,0}^{(0,-2)}$ | -2.85E-06 | 3.91E-03 |
| $\alpha_{1,0}^{(0,-1)}$ | 4.74E-03 | 7.20E-02 |
| $\alpha_{1,0}^{(0,0)}$ | 2.43E-02 | 2.36E-01 |
| $\alpha_{1,0}^{(0,1)}$ | 4.74E-03 | 7.20E-02 |
| $\alpha_{1,0}^{(0,2)}$ | -2.85E-06 | 3.91E-03 |
| $\alpha_{1,0}^{(0,3)}$ | 1.75E-09 | 2.30E-04 |

In Table 2 we present some results for two types of DLW – with phase velocity close to velocity of light ($\beta \sim 1$, case 1 and 2) and with small phase velocity ($\beta \sim 0.5$, case 3 and 4).

In all our calculations disc thickness was $t = 0.4$ cm.

Results were obtained for the number of waveguide modes $S_w = 50$ (see (1.9), $1 \le s \le S_w$ in systems (1.26) and (1.27)) and the number of resonator radial modes $M_R = 500$ (see (1.17), $1 \le m \le M_R$).

**Table 2**

|   | $a$, cm | $b$, cm | $d$, cm | $N$ | $n$ | $f_*$ (GHz) | $f$ (SFish) | $\varphi_*$ (°) | $\varphi$ |
|---|---|---|---|---|---|---|---|---|---|
| 1 | 1 | 4.075 | 3.1 | 4 |   | 2.8927965 | 2.8930081 | 120 |   |
|   |   |   |   |   | 3 |   |   |   | 120.000 |
|   |   |   |   |   | 2 |   |   |   | 119.979 |
|   |   |   |   |   | 1 |   |   |   | 120.559 |
| 2 | 1.8 | 4.075 | 3.1 | 4 |   | 3.0427781 | 3.0432933 | 120 |   |
|   |   |   |   |   | 3 |   |   |   | 119.9999 |
|   |   |   |   |   | 2 |   |   |   | 119.985 |
|   |   |   |   |   | 1 |   |   |   | 119.49 |
| 3 | 1 | 4.075 | 1.55 | 4 |   | 2.8927965 | 2.8930081 | 120 |   |
|   |   |   |   |   | 3 |   |   |   | 120.000 |
|   |   |   |   |   | 2 |   |   |   | 119.979 |
|   |   |   |   |   | 1 |   |   |   | 120.559 |
| 4 | 1.8 | 4.075 | 1.55 | 4 |   | 3.2243889 | 3.2250822 | 120 |   |
|   |   |   |   |   | 3 |   |   |   | 120.007 |
|   |   |   |   |   | 2 |   |   |   | 119.770 |
|   |   |   |   |   | 1 |   |   |   | 121.605 |

From Table 2 it follows that the case $N = 3$ can be used for simulation of almost all types of DLW that were developed for accelerator application. The truncation scheme for the case $N = 3$ is depicted in Fig. 3.

In Table 3 ($a = 1.8$ cm, $b = 4.075$ cm, $d = 1.55$ cm) and Table 4 ($a = 1$ cm, $b = 4.075$ cm, $d = 1.55$ cm) results are presented from which we can conclude that $S_w \ge 50$ and $M_R \ge 500$ are suitable values for the number of waveguide modes and the number of resonator radial modes. Let's note that such inequality have to be fulfilled: $M_R \gg S_w \dfrac{b}{a}$ (see (1.17)).

**Table 3**

| $S_w$ | $M_R$ | $f_*$ | $f$ (SuperFish) |
|---|---|---|---|
| 25 | 500 | 3.223963 | 3.225082 |
| 50 | 500 | 3.224388 |   |
| 75 | 500 | 3.224527 |   |
| 100 | 500 | 3.224567 |   |
| 125 | 500 | 3.224615 |   |
| 150 | 500 | 3.224636 |   |
|   |   |   |   |
| 50 | 500 | 3.224388 |   |
| 50 | 750 | 3.224399 |   |
| 50 | 1000 | 3.224381 |   |
| 50 | 1250 | 3.224396 |   |
| 50 | 1500 | 3.224396 |   |

**Table 4**

| $S_w$ | $M_R$ | $f_*$ | $f$ (SuperFish) |
|---|---|---|---|
| 50 | 500 | 2.892796 | 2.893008 |
| 100 | 500 | 2.892834 | |
| | | | |
| 50 | 500 | 2.892796 | |
| 50 | 1000 | 2.892792 | |

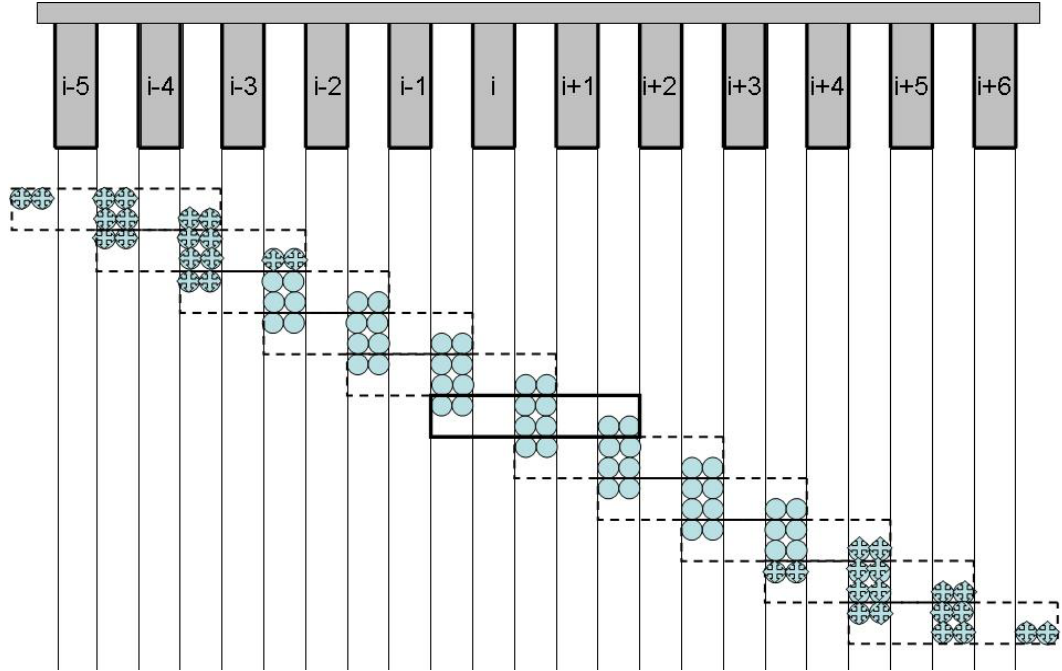

**Fig. 3 Truncation scheme for $N = 3$**

Analysis of the main coupling equations

$$\left(\omega_{1,0}^{(k)2} - \omega^2\right)e_{1,0}^{(k)} = \omega_{1,0}^{(k)2} \sum_{j=k-3}^{k+3} e_{1,0}^{(j)} \alpha_{1,0}^{(k,j)}. \qquad (1.38)$$

together with expression (1.29) and systems (1.26) and (1.27) shows that it is necessary to take into account coupling of fields (on the right and left sides) in (3+1+3)+3+3=13 resonators.

On the base of coupling equations we can study the processes of wave propagation in inhomogeneous DLWs. It can be done if we suppose that before and after the inhomogeneous DLW there are homogeneous fragments of DLW. In this case in homogeneous fragments of DLW at sufficient distance[3] from the connection interfaces (when all evanescent wave decay) we can search amplitudes in the form

$$e_{1,0}^{(k)} = \begin{cases} \exp\{i\varphi_{1,0}(k-k_1)\} + R\exp\{-i\varphi_{1,0}(k-k_1)\}, & k < k_1 \\ T\exp\{i\varphi_{2,0}(k-k_2)\} & k > k_2 \end{cases} \qquad (1.39)$$

and the first two equations in the coupling system will be such

---

[3] We must remember that the homogeneous fragments must contain more then 6 resonators.

$$\left(1-\frac{\omega^2}{\omega_{1,0}^{(k_1-1)2}}\right)\left[\exp\{-i\varphi_{1,0}\}+R\exp\{i\varphi_{1,0}\}\right]e_{1,0}^{(k_1-1)} = \sum_{j=k_1}^{k_1+2}e_{1,0}^{(j)}\alpha_{1,0}^{(k_1,j)} +$$
$$+\sum_{j=k_1-4}^{k_1-1}\alpha_{1,0}^{(k_1,j)}\left[\exp\{i\varphi_{1,0}(j-k_1)\}+R\exp\{-i\varphi_{1,0}(j-k_1)\}\right]$$
(1.40)

$$\left(1-\frac{\omega^2}{\omega_{1,0}^{(k_1)2}}\right)e_{1,0}^{(k_1)} = \sum_{j=k_1}^{k_1+3}e_{1,0}^{(j)}\alpha_{1,0}^{(k_1,j)} + \sum_{j=k_1-3}^{k_1-1}\alpha_{1,0}^{(k_1,j)}\left[\exp\{i\varphi_{1,0}(j-k_1)\}+R\exp\{-i\varphi_{1,0}(j-k_1)\}\right].$$
(1.41)

The equation (1.40) is the additional equation for determining the reflection coefficient $R$. The analogue equation for transition coefficient can be obtained from the equation (1.38) with $k=k_2+1$. So, our system will have $(k_2-k_1+3)$ unknown values ($e_{1,0}^{(k)}, k_1 \le k \le k_2, R, T$) and the same number of equations.

**Table 5**

| | | | | | | | |
|---|---|---|---|---|---|---|---|
| 1 | 1.4 | 4.16874 | 3.0989 | 1 | 1.4 | 4.16874 | 3.0989 |
| 2 | 1.4 | 4.16874 | 3.0989 | 2 | 1.4 | 4.16874 | 3.0989 |
| 3 | 1.4 | 4.16874 | 3.0989 | 3 | 1.4 | 4.16874 | 3.0989 |
| 4 | 1.4 | 4.16874 | 3.0989 | 4 | 1.4 | 4.16874 | 3.0989 |
| 5 | 1.4 | 4.16874 | 3.0989 | 5 | 1.4 | 4.16874 | 3.0989 |
| 6 | 1.4 | 4.16874 | 3.0989 | 6 | 1.4 | 4.16874 | 3.0989 |
| 7 | 1.4 | 4.16874 | 3.0989 | 7 | 1.4 | 4.16874 | 3.0989 |
| 8 | 1.4 | 4.16874 | 3.0989 | 8 | 1.4 | 4.16874 | 3.0989 |
| 9 | 1.4 | 4.16874 | 3.0989 | 9 | 1.4 | 4.16874 | 3.0989 |
| 10 | 1.4 | 4.16874 | 3.0989 | 10 | 1.4 | 4.16874 | 3.0989 |
| 11 | 1.4 | 4.16874 | 3.0989 | 11 | 1.4 | 4.16874 | 3.0989 |
| 12 | 1.4 | 4.16874 | 3.0989 | 12 | 1.4 | 4.16874 | 3.0989 |
| 13 | 1.4 | 4.16874 | 3.0989 | 13 | 1.4 | 4.16874 | 3.0989 |
| 14 | 1.4 | 4.16874 | 3.0989 | 14 | 1.4 | 4.16874 | 3.0989 |
| 15 | 1.4 | 4.16874 | 3.0989 | 15 | 1.4 | 4.16874 | 3.0989 |
| 16 | 1.3705 | 4.15053 | 3.0989 | 16 | 1.36905 | 4.19335 | 2.225 |
| 17 | 1.3 | 4.14066 | 3.0989 | 17 | 1.3 | 4.14066 | 3.0989 |
| 18 | 1.3 | 4.14066 | 3.0989 | 18 | 1.3 | 4.14066 | 3.0989 |
| 19 | 1.3 | 4.14066 | 3.0989 | 19 | 1.3 | 4.14066 | 3.0989 |
| 20 | 1.3 | 4.14066 | 3.0989 | 20 | 1.3 | 4.14066 | 3.0989 |
| 21 | 1.3 | 4.14066 | 3.0989 | 21 | 1.3 | 4.14066 | 3.0989 |
| 22 | 1.3 | 4.14066 | 3.0989 | 22 | 1.3 | 4.14066 | 3.0989 |
| 23 | 1.3 | 4.14066 | 3.0989 | 23 | 1.3 | 4.14066 | 3.0989 |
| 24 | 1.3 | 4.14066 | 3.0989 | 24 | 1.3 | 4.14066 | 3.0989 |
| 25 | 1.3 | 4.14066 | 3.0989 | 25 | 1.3 | 4.14066 | 3.0989 |
| 26 | 1.3 | 4.14066 | 3.0989 | 26 | 1.3 | 4.14066 | 3.0989 |
| 27 | 1.3 | | | 27 | 1.3 | | |

We studied the process of wave propagation in different inhomogeneous DLWs. Some results one can find in the preprint [12]. Here we present results of studying the possibility of matching two DLWs with the same phase shift per cell $\varphi = 2\pi/3$ but different aperture sizes. It is common for considering as the best cells for matching $\pi/2$ symmetric[4] cells (see, for example, [13]). This DLW construction cannot be simulated by described above method. So, we consider the possibility of matching two DLWs with nonsymmetric cells. Consideration was carried out for frequency $f = 2856$ MHz.

---

[4] Cells that have the same segments of circular waveguides from the both sides of each disc

The DLW represents the matched (on one frequency $f = 2856$ MHz) connection of two different uniform DLWs with using one transition cell with $d = 3.0989$ cm ($\omega(d+t)/c = 2\pi/3$) and $d = 2.225$ cm ($\omega(d+t)/c = \pi/2$). Parameters of DLWs are given in Table 5 (disc thickness $t = 0.4$ cm). Transition cell sizes (a(16) and b(16)) were chosen by making the reflection coefficient small enough ($|\Gamma| = 2.5 \times 10^{-4}$ and $|\Gamma| = 3.2 \times 10^{-4}$). Results of calculations that presented in Fig. 5 show that it is impossible to match two DLWs with nonsymmetric cells without additional phase shift. This additional phase shift. is smaller for $\pi/2$ nonsymmetric transion cell.

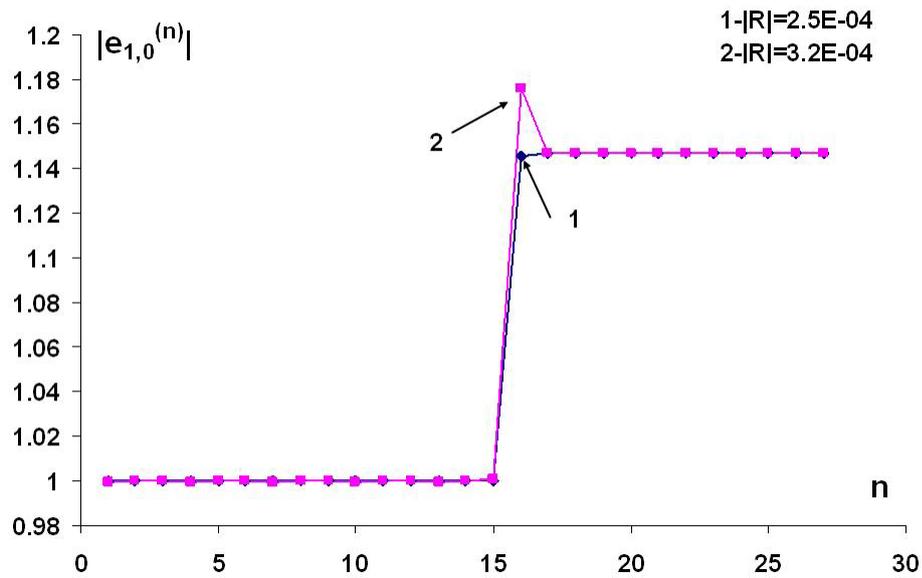

**Fig. 4** Amplitude distribution along the DLW: 1- $2\pi/3$ transition cell, 2- $\pi/2$ transition cell

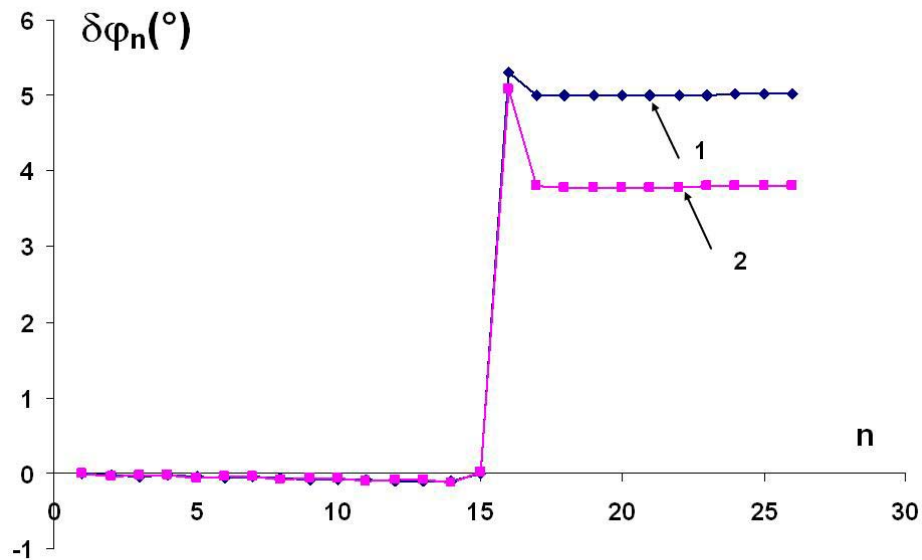

**Fig. 5** Distribution of differences $\delta\varphi_n = \varphi_n - 120°$ along the DLW: 1- $2\pi/3$ transition cell, 2- $\pi/2$ transition cell

**Conclusions**

Results of conducted investigations show that proposed mode matching technique can be effectively used for calculation of parameters of nonuniform DLWs. There is one problem in the developed coupled cavity model – taking into account the rounding of the disk hole edges [2].